\documentclass[floats,amssymb,nofootinbib,showkeys,twocolumn]{revtex4}
\usepackage{amssymb}
\usepackage{graphics}
\usepackage{amsmath}
\usepackage{float}
\usepackage{xcolor}
\usepackage{amsfonts}
\usepackage{bm}% bold math
\usepackage[]{latexsym}

\newcommand{\be}{\begin{equation}}\newcommand{\ee}{\end{equation}}
\newcommand{\bea}{\begin{eqnarray}}\newcommand{\eea}{\end{eqnarray}}
\newcommand{\brr}{\begin{array}}\newcommand{\err}{\end{array}}
\newcommand{\bit}{\begin{itemize}}\newcommand{\eit}{\end{itemize}}
\newcommand{\ben}{\begin{enumerate}}\newcommand{\een}{\end{enumerate}}

\newcommand{\ba}{\begin{array}}
\newcommand{\ea}{\end{array}}

\begin{document}
\title{Primordial Big Bang Nucleosynthesis and Generalized Uncertainty Principle}
%%%%%%%%%%%%%%%%%%%%%%%%%%%%%%%%%%%%%%%%%%%%%%%%%%%%%%%%%%%%%%%%%%%%%%%%%%%%%%%%%%%%%%%%%%%%%%%%%%%%%%%%%%%%%%%%%%%

\author{Giuseppe Gaetano Luciano\footnote{gluciano@sa.infn.it}$^{\hspace{0.3mm}1,2}$}
 
\affiliation
{$^1$Dipartimento di Fisica, Universit\`a di Salerno, Via Giovanni Paolo II, 132 I-84084 Fisciano (SA), Italy
\\ 
\vspace{0mm}
$^2$INFN, Sezione di Napoli, Gruppo collegato di Salerno, Italy}

\date{\today}

\begin{abstract}
The Generalized Uncertainty Principle (GUP) naturally
emerges in several quantum gravity models, predicting
the existence of a minimal length at Planck scale. 
Here, we consider the quadratic GUP as a semiclassical approach
to thermodynamic gravity and constrain the deformation parameter
by using observational bounds from Big Bang Nucleosynthesis and
primordial abundances of the light elements ${}^4 He, D, {}^7 Li$. We show that our result fits with most of existing bounds on $\beta$
derived from other cosmological studies.  

\end{abstract}


\date{\today}

\maketitle

\section{Introduction}
\label{Intro}
Quantum Theory and General Relativity are the two best 
descriptions of Nature to date. On one hand, Quantum Mechanics 
governs the properties of matter at microscopic scales, 
laying the foundations of solid state physics. By contrast, 
General Relativity deals with large-scale phenomena in the Cosmos -   
from the solar system to the faraway galaxies - as well as with the
evolution of the Universe as a whole. In spite of providing
successful predictions in their respective domains, 
these two theories exhibit fatal inconsistencies when combined together.
Much effort has been devoted to the construction of a unified formalism in the
last decades, culminated with the development of a number of promising candidate models.  Yet despite this striving, a definitive answer
is still far from being reached, thus making the quantization of gravity
a central open question in modern theoretical physics. 

A distinctive signature of most approaches to quantum gravity (QG) 
is the emergence of a minimal measurable length at around Planck energy. 
Implications of this fundamental scale are often taken into account
by deforming the Heisenberg Uncertainty Principle (HUP)~\cite{Amati,Konishi,Maggiore,Kempf,Scard,Capozzi,Adler,Mague}, so as
to accommodate a minimal uncertainty in position measurements.
The most common form of generalized uncertainty principle (GUP)
is obtained by adding a term quadratic in the momentum over
the standard Heisenberg limitation, i.e.
\be
\label{GUP}
\Delta x\,\Delta p \gtrsim \hbar\left[1+4\beta\left(\frac{\Delta p}{m_p\hspace{0.2mm}c}\right)^2 \right], 
\ee
where the pre-factor has been set of order unity, as seen in~\cite{Garay,Medved,AmeCamel}. Here, $m_p\simeq10^{19}\,\mathrm{GeV}$ 
denotes the Planck mass. To simplify the notation, henceforth we work
in natural units $\hbar=1=c$.

The (dimensionless) deformation parameter $\beta$ is
not fixed by the theory, leaving room for an intensive research activity~\cite{Brau,Das,Pedram,ScardCas,QC,Petroz,CGgrav,Buoninf,AliTest,Bruk,GravBar,Bawaj,BossoLigo,Pendu,LucLuc,LucianoCas,Kana,BossoLuc} (see Table~\ref{Tab1} and Table~\ref{Tab2} for upper bounds of cosmological and quantum/gravitational origin, respectively). 
Debate also concerns the sign of $\beta$: although it is assumed to be positive in the original formulation of the GUP, arguments   
in favor of negative values are not missing~\cite{Mague,JizbaKl,Ong,CGgrav}. 

One of the contexts in which the GUP has been studied
most extensively is that of black holes (BH's). In particular, in~\cite{Adler} 
it has been shown that Eq.~\eqref{GUP} inevitably affects Hawking
temperature and the related BH evaporation process, with a non-trivial impact on the whole BH thermodynamics. Likewise, GUP-induced 
corrections enter the Bekenstein-Hawking entropy formula, resulting in a generalized Bekenstein bound~\cite{BLPS} and a modified area law~\cite{Medved,Anacleto}. Remarkably, implications of the 
modified area law are also explored at cosmological level, because of the geometrical - and therefore universal - nature of this law, which can be applied to any causal horizon~\cite{Jacob}. 

The tight interweaving of BH horizon thermodynamics
and GUP has renewed the interest for \emph{thermodynamic gravity}.
In this approach, Einstein field equations are derived from the first law of thermodynamics, combined with the entropy area law~\cite{Jacob}. 
An interesting consequence of this achievement is that one can recover 
the cosmological Friedmann equations by applying the
first law of thermodynamics to the apparent
horizon of the Friedmann-Lema\^itre-Robertson-Walker (FLRW)
spacetime~\cite{Wang,Frolov,Padmanabhan1,Eling,Akbar1}. 
This procedure has recently been proven to be quite general, 
being equally applicable in theories of gravity beyond General Relativity~\cite{Cai} and even in the presence of a modified entropy-area law~\cite{CaiCao}.
Along this line, in~\cite{Zhu} Friedmann equations
have been derived from the GUP-modified expression of the entropy, obtaining generalized (i.e. $\beta$-dependent) relations.
This indicates that GUP effects at high energies can
affect the dynamics of the FLRW Universe
at early times, albeit in a mild way. The resulting framework is often referred to as GUP Cosmology.

\begin{table}[t]
  \centering
    \begin{tabular}{c c c}
    \hline
    $|\beta|\lesssim$\hspace{0.9cm} & Physical framework\hspace{1.5cm} & Refs. \\
        \hline
             \hspace{-10.2mm}$10^{8}$ &\hspace{-3.46cm} Baryogenesis \hspace{-0.98cm}  & \cite{LambBar} \\[1mm]
                         \hspace{-9mm}$10^{59}$ &\hspace{-2.46cm} Full data Cosmology\hspace{-0.98cm}  & \cite{Giardino} \\[1mm]  
                          \hspace{-9mm}$10^{81}$ &\hspace{-2.43cm} ${}^4He$, $D$ Abundances
                          \hspace{-0.98cm} & [This work] \\[1mm]  
        \hspace{-9mm}$10^{81}$ &\hspace{-2.63cm} Type Ia supernovae\hspace{-0.98cm}  & \cite{Kouwn} \\[1mm]
                \hspace{-9mm}$10^{81}$ &\hspace{-1.49cm} Baryon acoustic oscillations\hspace{-0.98cm}  & \cite{Kouwn} \\[1mm]
                   \hspace{-9mm}$10^{81}$ &\hspace{-2.4cm} Late-time Cosmology\hspace{-0.98cm}  & \cite{Giardino} \\[1mm]  
                               \hspace{-9mm}$10^{82}$ &\hspace{-3.2cm} ${}^7Li$ Abundance
\hspace{-0.98cm}  & [This work]\\[1mm]  
                        \hspace{-9mm}$10^{87}$ &\hspace{-2.1cm} Freeze-out temperature\hspace{-0.98cm}  & [This work]\\
                          \hline
            \end{tabular}
  \caption{Upper bounds on the GUP parameter from cosmological analysis.}
  \label{Tab1}
\end{table}

\begin{table}[t]
  \centering
    \begin{tabular}{c c c}
    \hline
    $\hspace{-0.2mm}|\beta|\lesssim$\hspace{0.9cm} & \hspace{-0.25cm}Physical framework\hspace{1.5cm} & Refs. \\
        \hline
                \hspace{-10.7mm}$10^{6}$ &\hspace{-2.58cm} Harmonic oscillators \hspace{-0.98cm}  & \cite{Pendu} \\[1mm]  
                    \hspace{-9.65mm}$10^{21}$ &\hspace{-1.08cm} Scanning tunneling microscope \hspace{-0.98cm}  & \cite{Das} \\[1mm]      
                                  \hspace{-9.65mm}$10^{21}$ &\hspace{-1.95cm} Equiv. princip.  violation \hspace{-0.98cm}  & \cite{Ghosh} \\[1mm]    
                       \hspace{-9.65mm}$10^{27}$ &\hspace{-1.13cm} Weak equiv. princip. violation \hspace{-0.98cm}  & \cite{Gao} \\[1mm]                               
                                      \hspace{-9.65mm}$10^{33}$ &\hspace{-2.41cm} Gravity bar detectors \hspace{-0.98cm}  & \cite{GravBar} \\[1mm]                                               
                 \hspace{-9.65mm}$10^{36}$ &\hspace{-3.95cm} Lamb shift \hspace{-0.98cm}  & \cite{Das,AliTest} \\[1mm]      
                                   \hspace{-9.65mm}$10^{36}$ &\hspace{-1.56cm} Interferometry experiments \hspace{-0.98cm}  & \cite{Luciano:2021cna} \\[1mm]          
                                         \hspace{-9.65mm}$10^{39}$ &\hspace{-1.52cm} $^{87}Rb$ Cold atom experiment \hspace{-0.98cm}  & \cite{ColdGao} \\[1mm]      
                                           \hspace{-9.65mm}$10^{50}$ &\hspace{-3.52cm} Landau levels \hspace{-0.98cm}  & \cite{Das} \\[1mm]  
                                             \hspace{-9.65mm}$10^{60}$ &\hspace{-2.63cm} Gravitational waves \hspace{-0.98cm}  & \cite{Feng} \\[1mm]       
                                              \hspace{-9.65mm}$10^{69}$ &\hspace{-2.5cm} Perihelion precession \hspace{-0.98cm}  & \cite{ScardCas} \\[1mm]       
                       \hspace{-9.65mm}$10^{71}$ &\hspace{-2.24cm} Pulsar periastron shift \hspace{-0.98cm}  & \cite{ScardCas} \\[1mm]           
                           \hspace{-9.65mm}$10^{72}$ &\hspace{-2.69cm} Geodetic precession \hspace{-0.98cm}  & \cite{Shap} \\[1mm]     
                                                                        \hspace{-9.65mm}$10^{73}$ &\hspace{-2.29cm} Gravitational red-shift \hspace{-0.98cm}  & \cite{Shap} \\[1mm] 
                                                                         \hspace{-9.65mm}$10^{77}$ &\hspace{-1.88cm} Quasiperiodic oscillations \hspace{-0.98cm}  & \cite{Jusufi} \\[1mm]
                                                                                    \hspace{-9.65mm}$10^{78}$ &\hspace{-3.2cm} Light deflection \hspace{-0.98cm}  & \cite{ScardCas} \\[1mm]  
                                   \hspace{-9.65mm}$10^{78}$ &\hspace{-2.8cm} Shapiro time delay \hspace{-0.98cm}  & \cite{Shap} \\[1mm]     
                                            \hspace{-10mm}$10^{90}$ &\hspace{-2.838cm} BH shadow (M87*)\hspace{-0.98cm}  & \cite{Neves} \\[1mm]                                                                                                     
           \hline
    \end{tabular}
  \caption{Upper bounds on the GUP parameter from quantum and gravitational experiments.}
  \label{Tab2}
\end{table}

Besides the plethora of theoretical studies
on the GUP, a research direction widely pursued 
in QG phenomenology is attempting to quantify the 
magnitude of GUP corrections by constraining
the deformation parameter. This is particularly useful 
in that it paves the way for a low-energy investigation
of QG, which could be somehow interfaced with experimental data. 
Nevertheless, to the best of our knowledge, 
situations where this kind of analysis is performed 
in GUP Cosmology are quite rare in the literature, 
as witnessed by the low number of bounds listed in Table~\ref{Tab1}. 
If on one hand this can be understood by observing
that bounds of cosmological origin are less stringent than
those obtained through quantum/gravitational experiments, 
on the other hand it should be acknowledged that these bounds  
can be derived with very high precision, due to the great
and accurate amount of cosmological data available to date. 

Starting from the above premises, the aim of this work
is to explore the implications of GUP Cosmology on 
Big Bang Nucleosynthesis (BBN).  
BBN describes the sequence of nuclear
reactions responsible for the synthesis of primordial light elements, such as Hydrogen $H$, its isotope Deuterium $D$, Helium isotopes ${}^3 He$ and ${}^4 He$
and Lithium isotope ${}^7 Li$~\cite{Kolb,Bern,PDG}. 
It is believed to have taken place shortly after the Big Bang, 
when the Universe was cooled enough to form stable protons and neutrons. 
Since BBN drives the observed Universe, 
it is clear that primordial abundances must be very tightly constrained in order
to reproduce the current chemical composition of the Universe.  
This fact promotes BBN as one of the best arena to constrain cosmological models.
In particular, in what follows we shall fix the GUP
parameter by requiring consistency between GUP Cosmology predictions and  
\emph{i}) the existing upper bound on the variations of the freeze-out
temperature, \emph{ii}) the current estimates of the primordial abundances
of ${}^4 He$, $D$ and ${}^7 Li$. We show that the ensuing upper bound on $\beta$ is consistent with most of existing constraints  
derived from other cosmological analysis. The results here
discussed could contribute to the debate of fixing 
the most reliable scenario among cosmological models
based on the GUP and also provide a possible explanation
for the ${}^7 Li$ puzzle.

The layout of the paper is as follows: in Sec.~\ref{MFE} we
review the derivation of the modified Friedmann equations within 
GUP framework. Toward this end, we follow~\cite{AliGUP,Ghos}. In Sec.~\ref{BBN} and Sec.~\ref{PRIM} we constrain the GUP parameter based on  observational data from BBN and primordial abundances, respectively. Section~\ref{DC} is devoted to conclusions and outlook. 
 
\section{Modified Friedmann equations from GUP}
\label{MFE}
In this Section we summarize the main steps leading
to the cosmological Friedmann equations and their
generalization to the GUP framework. As usual, we assume
that, for a homogeneous and isotropic $(1+3)$-dimensional FRW Universe, 
the line element is given by
\be
ds^2=h_{ab}dx^{a}dx^{b}+\tilde r^2(d\theta^2+\sin^2\theta d\phi^2)\,,\,\,\,\, a,b=\{0,1\},
\ee
where $h_{ab}=\mathrm{diag}(-1,a^2/(1-kr^2))$ is the metric
of a $(1+1)$-dimensional subspace, $x^a=(t,r)$,
$\tilde r=a(t)r$, with $a(t)$ being the time-dependent 
scale factor, $r$ is the comoving
radius and $k$ the (constant) spatial curvature.  $\theta,\phi$
are the angular coordinates. 

One can think of the Universe as a physically bounded region
of (apparent) horizon radius
\be
\label{rtilde}
\tilde r_A=\frac{1}{\sqrt{H^2+\frac{k}{a^2}}}\,,
\ee
and temperature
\be
\label{T}
T=-\frac{1}{2\pi \tilde r_A}\left(1-\frac{\dot{\tilde r}_A}{2H\tilde r_A}\right), 
\ee
where $H=\dot a(t)/a(t)$ is the Hubble parameter (the dot denotes time derivative). 
For our later purposes, we can roughly neglect
the  space curvature $k$, so that Eq.~\eqref{rtilde} 
reads $\tilde r_A\simeq1/H$.

By describing the 
matter and energy content of the Universe as a perfect fluid, the
energy-momentum tensor is
\be
T_{\mu\nu}=(\rho+p)u_\mu u_{\nu}+pg_{\mu\nu}\,,
\ee 
where $u_{\mu}$, $\rho$ and $p$ are the four-velocity, 
energy density and pressure of the fluid, respectively. 
The continuity equation 
\be
\label{coe}
\dot \rho=-3H\left(\rho+p\right)\,,
\ee 
holds true. 

Based on the deep connection between 
gravity and thermodynamics~\cite{Jacob}, the Friedmann equations
in the bulk of the Universe follow from the first law of thermodynamics
\be
\label{FLT}
dE=TdS+WdV\,,
\ee
applied on the boundary. Here, the total energy of the matter
existing inside the apparent horizon of entropy $S$ is given by $E=\rho V$, 
with $V=4\pi\hspace{0.2mm}\tilde r_A^3/3$ being the volume enclosed by the horizon. 
The work density $W$ is related to the
energy density and pressure   
by $W=-\frac{1}{2}T^{ab}h_{ab}=\frac{1}{2}(\rho-p)$.

In standard Cosmology the horizon entropy obeys
the holographic principle 
\be
\label{holo}
S=\frac{A}{4G}\,,
\ee
where $A=4\pi\hspace{0.2mm}\tilde r_A^2$ is the horizon surface area ($G$ denotes Newton's 
gravitational constant). With this as physical input, 
it is a straightforward text-book exercise to show that
Eq.~\eqref{FLT} leads to the Friedmann equations for a flat Universe
\begin{eqnarray}
\label{FE1}
H^2&=&\frac{8}{3}\hspace{0mm}\pi\hspace{0.2mm} G\rho\,,\\[2mm]
\dot H&=&-4\pi\hspace{0.2mm} G\left(\rho+p\right)\,.
\label{FE2}
\end{eqnarray}

Following~\cite{AliGUP}, we now suppose
that the general expression for the GUP-modified 
entropy-area law takes the form
\begin{eqnarray}
\label{mod}
S&=&\frac{f(A)}{4G}\,,\\[2mm]
\frac{dS}{dA}&=&\frac{f'(A)}{4G}\,,
\label{dsda}
\end{eqnarray}
where the function $f(A)$ is to be determined ($f'(A)$ denotes
the derivative of $f$ respect to $A$). 
For the quadratic GUP model~\eqref{GUP}, 
this can be done by computing the minimal change 
of area $\Delta A_{min}= 8\pi\ell_p^2\hspace{0.2mm} E\hspace{0.2mm}\Delta x$ of an apparent horizon absorbing a quantum
particle of given energy $E\simeq\Delta p$ and finite size $\Delta x\simeq r_s=\sqrt{A/\pi}$ ($r_s=2\hspace{0.2mm}M\hspace{0.2mm}G$ is the Schwarzschild radius).  
After some algebra, one gets~\cite{AliGUP,LambBar}
\be
\frac{dS}{dA}=\frac{\Delta S_{min}}{\Delta A_{min}}=\frac{1+\sqrt{1-\beta^*/A}}{8\ell_p^2}\,,
\ee
where $\Delta S_{min}=\ln2$ is the minimal increase in entropy, corresponding to one bit of information. Here, we have defined $\beta^*\equiv16\pi \hspace{0.2mm}\beta\hspace{0.2mm}\ell_p^2$
and $\ell_p=1/m_p=\sqrt{G}$ is the Planck length.
Comparison with Eq.~\eqref{dsda} allows us to identify
\begin{equation}
\label{fpA}
f'(A)=\frac{1+\sqrt{1-\beta^*/A}}{2}\,. 
\end{equation}
It is easy to check that $f'(A)\rightarrow1$ for vanishing $\beta^*$, 
consistently with the holographic relation~\eqref{holo}. 
By plugging Eq.~\eqref{fpA} into~\eqref{dsda}, 
and integrating over $A$, it is also possible to derive
the explicit formula for the 
GUP-modified Bekenstein-Hawking entropy. The resulting
expression is rather awkward to exhibit. Since
we do not need it explicitly in the following analysis, we 
remand the interested reader to~\cite{AliGUP,LambBar}. 

We have now all the ingredients to infer GUP effects
on Friedmann equations. Indeed, by replacing Eq.~\eqref{dsda} and~\eqref{fpA} into the first law of thermodynamics~\eqref{FLT} and noticing that
\be
dE=4\pi\rho\hspace{0,2mm}\tilde r_A^2d\tilde r_A+\frac{4}{3}\pi \tilde r_A^3\hspace{0.2mm}d\rho
\ee 
on the horizon surface, we are led to
\be
\label{ModF1}
\frac{4\pi}{\tilde r_A^3}\left(1+\sqrt{1-\frac{\beta^*}{4\pi\tilde r_A^2}}\right)d\tilde r_A=-\frac{32}{3}\pi^2\hspace{0.2mm}G\,d\rho\,.
\ee
After integrating the l.h.s. between $\tilde r_A$ and the minimal length-scale $\Delta x_{min}\simeq\sqrt{\beta^*/\pi}$ allowed by the GUP~\eqref{GUP} 
and setting the integration constant $\rho\left(\Delta x_{min}\right)$ in such a way that 
Eq.~\eqref{FE1} is recovered for $\beta^*\rightarrow0$, we obtain to the leading order in the deformation parameter\footnote{Strictly speaking, we are expanding around $\epsilon\equiv \beta\hspace{0.2mm}\ell_p^4\hspace{0.2mm}\rho$. We shall check \emph{a posteriori} the degree of validity of this approximation (see Sec.~\ref{BBN}).}
\be
\label{Hsi}
H_\beta(\rho)=H(\rho)\left(1+\frac{2\beta}{3}\hspace{0.2mm}\pi\hspace{0.2mm}G^2\rho\right),
\ee
with $H$ being the standard Hubble parameter given by Eq.~\eqref{FE1}. 
This relation provides the first GUP-modified Friedmann equation. 
For later convenience, we recast it in the form
\be
\label{HsiZ}
H_\beta(\rho)=H(\rho)\hspace{0.2mm}Z_\beta(\rho)\,,
\ee
where we have separated out the $\beta$-dependence of $H_\beta$ by defining
\be
\label{Z}
Z_\beta(\rho)=1+\frac{2\beta}{3}\hspace{0.2mm}\pi\hspace{0.2mm}G^2\rho\,.
\ee
In view of applying the above formalism to BBN,  
we can further manipulate Eq.~\eqref{HsiZ} by using the relation
\be
\label{rhoT}
\rho=\frac{\pi^2\hspace{0.2mm} g(T)}{30}\,T^4\,,
\ee
where $g(T)$ denotes the effective number of degrees
of freedom. Equation~\eqref{HsiZ} becomes
\be
\label{HT}
H_\beta(T)=H(T)\hspace{0.2mm}Z_\beta(T)\,,
\ee
where
\begin{eqnarray}
\label{HTbis}
H(T)&=&\frac{2\pi}{3}\sqrt{\frac{\pi\hspace{0.3mm}G\hspace{0.3mm} g(T)}{5}}\hspace{0.2mm}T^2\,,\\[2mm]
Z_\beta(T)&=&1+\frac{\beta}{45}\hspace{0.2mm}\pi^3\hspace{0.3mm}G^2\hspace{0.2mm}g(T)\hspace{0.3mm} T^4\,.
\label{Zbetabis}
\end{eqnarray}

In a similar fashion, one can derive the linearized second GUP-modified
Friedmann equation to be~\cite{LambBar}
\be
\label{SGFM}
\dot{H}_\beta=\dot H\left(1+\beta\hspace{0.2mm}G\hspace{0.2mm} H^2\right),
\ee
which still recovers Eq.~\eqref{FE2} in the limit of vanishing $\beta$.

These GUP-corrected Friedmann equations form the basis
on which variations of the Hubble parameter and of its time derivative 
in the early Universe will be studied.

\section{Big Bang Nucleosynthesis in GUP Cosmology}
\label{BBN}
In this Section we study the BBN within the framework
of GUP Cosmology. We assume that the energy density
of relativistic particles filling up the Universe is given by Eq.~\eqref{rhoT}
with $g(T)=g_*\simeq10$ (henceforth we consider the radiation dominated
era), the major contribution to the degrees of freedom being given by relativistic photons, $e^+e^-$ pairs
and the three neutrino species.. 

According to the standard BBN model, neutron and protons
started to form only few thousandths of a second after the 
Big Bang, when the temperature dropped low enough. 
From the first hundredth of a second up to few minutes, 
the abundances of the first very light atomic nuclei were defined.
In particular, the formation of the primordial ${}^4 He$ took place at around $T\simeq 100\,\mathrm{MeV}$, 
while the energy and number density were still dominated by
relativistic leptons (electrons, positrons and neutrinos)
and photons. Due to their rapid collisions, such particles
were in thermal equilibrium, so that 
$T_\nu=T_e=T_\gamma=T$~\cite{Bern}. On the other hand, 
the smattering of 
protons and neutrons were kept in equilibrium 
owing to the following weak interactions with leptons
\begin{eqnarray}
\label{nu1}
&\emph{a)}\,\,\,\,\,\nu_e+n\,\, \longleftrightarrow\,\, p + e^{-}\,,\\[2mm]
\label{nu2}
&\emph{b)}\,\,\,\,\,e^{+}+n \,\,\longleftrightarrow\,\, p+\bar\nu_e\,,\\[2mm]
\label{nu3}
&\emph{c)}\,\,\,\,\,n \,\,\longleftrightarrow\,\, p+e^{-}+\bar\nu_e\,.
\end{eqnarray}

Within the framework outlined above, neutron abundance can be computed 
by estimating the conversion rate $\lambda_{pn}(T)$ of protons into neutrons and its inverse $\lambda_{np}(T)=e^{-\mathcal{Q}/T}\lambda_{pn}(T)$, where
$\mathcal{Q}=m_n-m_p\simeq 1.29\,\mathrm{MeV}$ is the difference between neutron and proton masses.
Here $\lambda_{np}$ is expressed
as the sum of the rates associated to the three processes~\eqref{nu1}-\eqref{nu3} separately, i.e.\footnote{Notice that the integration over momentum appearing in the definition of $\lambda_{a}$, $\lambda_{b}$ and $\lambda_{c}$ might be affected in the GUP framework due to minimal-length effects. However, we expect these corrections not to spoil significantly the order of magnitude of the resulting rates, thus being negligible in first approximation.} 
\be
\label{lnp}
\lambda_{np}(T)= \lambda_{a}(T)+\lambda_{b}(T)+\lambda_{c}(T)\,.
\ee 
In turn, the total weak interaction rate reads $\Lambda(T)=\lambda_{np}(T)+\lambda_{pn}(T)$.

Following~\cite{Bern}, we further assume that, during the freeze-out period, 
the temperature $T$ is low in comparison with the 
with the characteristic energies contributing to the rates for the decays~\eqref{nu1}-\eqref{nu3}. This allows us to estimate the 
lepton phase-space density functions by
the ''classical'' Boltzmann weights, rather than
the Fermi-Dirac distribution. The last requirement is that
the electron mass $m_e$ can be neglected with respect to the electron and neutrino energies. Under these conditions, one can show that~\cite{Bern,Ghos}
\be
\label{lalb}
\lambda_{a}(T)\simeq q T^5+\mathcal{O}\left({\frac{\mathcal{Q}}{T}}\right)=\lambda_{b}(T)\,,
\ee
where $q\simeq10^{-10}\,\mathrm{GeV}^{-4}$. 
On the other hand, the contribution of the free-neutron decay process $c)$
to the total rate is found to be negligible\footnote{For $T\gtrsim 1\,\mathrm{MeV}$
this contribution is 3 orders of magnitude lower than the rate~\eqref{lalb}~\cite{Bern}.}, implying that the total rate $\lambda_{np}(T)$ is roughly twice
that given in Eq.~\eqref{lalb}. 

The ${}^4 He$ mass fraction of the total baryonic mass is now estimated as~\cite{Kolb,Ghos}
\be
\label{yp}
Y_p\equiv \gamma \frac{2\hspace{0.2mm}x\left(t_f\right)}{1+x\left(t_f\right)}\,,
\ee
where $\gamma=e^{-(t_n-t_f)/\tau}\simeq 1$ depends on the (relatively short)
time between freeze-out ($t_f$) and nucleosynthesis ($t_n$)
and on the neutron mean lifetime $\tau\simeq877\,\mathrm{s}$.
It can be as the fraction of neutrons
that decay into protons in the interval $t\in[t_f,t_n]$.
$x(t_f)=e^{-\mathcal{Q}/T(t_f)}$ is the neutron-to-proton
equilibrium ratio. 

Deviations from $Y_p$ due to the variation of the
freeze-out temperature $T_f$ can be quantified as~\cite{Ghos}
\be
\label{dyp}
\delta Y_p=Y_p\left[\left(1-\frac{Y_p}{2\gamma}\right)\log\left(\frac{2\gamma}{Y_p}-1\right)-\frac{2t_f}{\tau}\right]\frac{\delta T_f}{T_f}\,,
\ee
where $\delta T_n$ has been set to zero, since $T_n$
is fixed by the $D$-binding energy~\cite{LambJCAP,CapLamb}. 

The mass fraction of ${}^4 He$ has been recently 
determined to a high degree of precision by making use
of infrared and visible ${}^4 He$ emission lines in extragalactic
HII regions, obtaining~\cite{Aver}
\be
\label{ypest}
Y_p=0.2449\,, \quad\,\, |\delta Y_p|\lesssim10^{-4}\,.
\ee
Insertion of these values into Eq.~\eqref{dyp} gives
\be
\label{dtsutf}
\left|\frac{\delta T_f}{T_f}\right|\lesssim10^{-4}\,,
\ee
where we have set $t_f\simeq 1\,\mathrm{s}$ and $t_n\simeq 20\,\mathrm{s}$. 

Following~\cite{Ghos}, we can compute the GUP-modified freeze-out
temperature $T_f$ by equating Eqs.~\eqref{HT}
and~\eqref{lnp}. 
With the further definition $\delta T_f=T_f-T_{0f}$, 
where $T_{0f}\simeq 0.6\, \mathrm{MeV}$~\cite{Ghos}, 
we get 
\be
\label{fTem}
\left|\frac{\delta T_f}{T_f}\right|=\left| 1-\frac{2\hspace{0.2mm}\beta\hspace{0.3mm}\pi^4\hspace{0.3mm}G^2\hspace{0.1mm}g_*\hspace{0.3mm}\sqrt{\pi\hspace{0.2mm} G\hspace{0.2mm} g_*/5}}{135\hspace{0.3mm}q}\hspace{0.3mm}T_{0f}
\right|\,.
\ee

The GUP parameter can be fixed
by demanding consistency between Eqs.~\eqref{dtsutf}
and~\eqref{fTem}. A straightforward numerical evaluation
leads to
\be
\label{firstest}
\beta\sim\mathcal{O}(10^{87})\,.
\ee
This means that our linearized approximation
is well-justified, since for this value of $\beta$ we have $\epsilon\sim\mathcal{O}(10^{-2})$ (see footnote 1). 

By comparison with bounds in Table~\ref{Tab1}, 
we see that the result~\eqref{firstest}
provides us with a weak cosmological constraint on $\beta$. 
The gap becomes even wider if compared with bounds from gravitational/quantum experiments
(see Table~\ref{Tab2}), thus emphasizing the quite negligible r\^ole of GUP on cosmic scales.

\section{Primordial ${}^4 He, D, {}^7 Li$ Abundances in GUP Cosmology}
\label{PRIM}

Let us now constrain the GUP by a slightly different approach.
The basic idea is to study GUP-induced
deviations from standard Cosmology on
the primordial abundances of Helium isotope ${}^4 He$, Deuterium $D$ and  Lithium isotope ${}^7 Li$. 
This will be done by replacing the standard
$Z$-factor entering primordial abundances
with the $\beta$-dependent $Z$-factor appearing in Eq.~\eqref{HT}. 

In this regard, we observe that in the ordinary Cosmology based on 
General Relativity, one simply has $Z=1$. Deviations of $Z$ from unity may arise
due to either modified descriptions of gravity
or the presence of additional light particles such as neutrinos,
in which case one has~\cite{Boran}
\be
\label{Zv}
Z_\nu=\left[1+\frac{7}{43}\left(N_\nu-3\right)\right]^{1/2}\,,
\ee
where $N_\nu$ is the number of neutrino generations. 
However, since we aim to focus on the effects of the
GUP on BBN, hereafter we assume $N_\nu=3$, ruling out the possibility
that in our framework departures of $Z$ from unity are originated by
degrees of freedom of additional particles. Given
the very tight observational constraints on the allowed primordial abundances, 
we expect in this way to infer reliable bounds on the deformation
parameter of GUP. 

\subsection{${}^4 He$ abundance}
In order to estimate ${}^4 He$ primordial abundance, we follow  
the approach of~\cite{Bhatta}, recently revived in~\cite{Ghos}. 
Let us summarize here the sequence of nuclear
reactions responsible for the production of this element. 
The first step consists in generating deuterium $D$ from a neutron
and a proton. After that, deuterium is converted  into ${}^3 He$ 
and tritium $T$. In short
\begin{eqnarray}
\label{D}
n+p&\rightarrow& D +\gamma\,,\\[2mm]
D+D&\rightarrow&{}^3 He+n\,,\\[2mm]
D+D&\rightarrow&T+p\,.
\end{eqnarray}
The last step of the chain leads to the production
of ${}^4 He$ due to the following processes
\begin{eqnarray}
D+T&\rightarrow&{}^4 He+n\,,\\[2mm]
D+{}^3 He&\rightarrow&{}^4 He+p\,.
\end{eqnarray}

According to~\cite{Kneller,Steig}, 
the numerical best fit constrains the primordial
${}^4 He$ abundance to be
\be
\label{ypb}
Y_p=0.2485 \pm 0.0006+0.0016\left[\left(\eta_{10}-6\right)+100\left(Z-1\right)\right],
\ee
where in our case we have to set $Z= Z_\beta$ given by Eq.~\eqref{Zbetabis}. 
Here, we have adopted the usual definition of the baryon
density parameter~\cite{Kneller,Steig} 
\be
\label{bdp}
\eta_{10}\equiv 10^{10}\eta_B\simeq6\,,
\ee
where $\eta_B$ is the 
baryon to photon ratio. Notice that, by setting $Z=1$, we recover the standard
${}^4 He$ abundance $Y_p=0.2485 \pm 0.0006$
predicted by BBN model.  

Now, as discussed in~\cite{Bhatta,Ghos}, consistency
between observational data on ${}^4 He$ abundance 
and Eq.~\eqref{ypb} with $\eta_{10}=6$
allows us to fix~\cite{Fields}
\be
\label{ZbHe}
\delta Z\equiv Z-1\lesssim\mathcal{O}(10^{-2})\,.
\ee
By using the expression~\eqref{Zbetabis}
for $Z_\beta$, we then obtain
\be
\label{bHE}
\beta\lesssim\mathcal{O}(10^{81})\,,
\ee
assuming $T\simeq10\,\mathrm{MeV}$
and
\be
\label{bHE2}
\beta\lesssim\mathcal{O}(10^{89})\,,
\ee
assuming $T\simeq0.1\,\mathrm{MeV}$.

Let us focus on the most stringent bound~\eqref{bHE}.
Except for the constraint of~\cite{LambBar} (which is however computed by referring to the much earlier
baryogenesis epoch\footnote{We point out that the
gap between the bound on $\beta$ from GUP baryogenesis~\cite{LambBar} and other cosmological bounds from different stages
of the evolution of the Universe
could be a hint for the need of a
GUP model with a time (or equivalently energy) dependent deformation parameter. Of course, such a running behavior might not be described through a simply (i.e. monotonically) decreasing function of time, but rather by a more complicated function. And indeed this should be the case in order to cure the above inconsistency. In this regard, we mention that a similar time-dependence of the deformed commutator occurs in Maguejo-Smolin Doubly Special Relativity~\cite{Mague}, which predicts that the generalized commutator should vanish at Planck scale, while approaching the conventional HUP at low energies.}) and that of~\cite{Giardino}
with full data Cosmology (which seems in general to display some inconsistencies between GUP and current data available on dark energy), 
we notice that the result $\beta\lesssim\mathcal{O}(10^{81})$
perfectly fits with other cosmological bounds
obtained via Type Ia supernovae~\cite{Kouwn} 
and baryon acoustic oscillations measurements~\cite{Kouwn} (see Table~\ref{Tab1}). 
It also agrees with late-time observational data from 
Early-Type Galaxies as Cosmic Chronometers,  
the H0 Lenses in COSMOGRAIL's Wellspring, the
``Mayflower'' sample of Gamma Ray Bursts and the
latest Planck 2018 release for Cosmic Microwave Background
radiation~\cite{Giardino}. %Furthermore, Eq.~\eqref{bHE} provides
%us with a more stringent bound than Eq.~\eqref{firstest}.

Again, we can see that
for the values of $\beta$ in Eqs.~\eqref{bHE},~\eqref{bHE2}, 
the approximation $\epsilon\ll1$ works well, since we have
$\epsilon\sim\mathcal{O}(10^{-3})$. 

\subsection{$D$ abundance}
Deuterium $D$ is generated form the process~\eqref{D}.
Following the same analysis as above, $D$ primordial abundance can 
be ascertained from the numerical best fit of~\cite{Steig2}, giving 
\be
\label{Dab}
y_{D_p}=2.6\left(1\pm0.06\right)\left(\frac{6}{\eta_{10}-6\left(Z-1\right)}\right)^{1.6}\,.
\ee
As before, the values $\eta_{10}=6$ and $Z=1$
yield the standard BBN prediction $y_{D_p}=2.6\pm0.16$. 

Observational constraints on $D$ abundance combined with 
Eq.~\eqref{Dab} allow us to set $\delta Z\lesssim\mathcal{O}(10^{-2})$~\cite{Bhatta,Ghos,Fields}, which is consistent with the constraint from ${}^2He$
abundance (see Eq.~\eqref{ZbHe}). 
Therefore, one still gets the bounds~\eqref{bHE}-\eqref{bHE2}
for $T\simeq (0.1 \div 10) \,\mathrm{MeV}$.

\subsection{${}^7 Li$ abundance}
It is well-known that 
the $\eta_{10}$ parameter which successfully fits the abundances
of ${}^4 He$, $D$  and other light elements is somehow inconsistent with 
observations of ${}^7 Li$. In fact, the ratio 
of the predicted value of ${}^7 Li$ abundance to the observed one 
lies in the interval $[2.4, 4.3]$ according to the standard cosmological theory~\cite{Boran,Fields2}. Quite unexpectedly, 
neither BBN nor any alternative model
are able to fit this so low abundance ratio. 
This puzzle is referred to as \emph{Lithium problem}. 

Once more, we can constrain deviations of $Z$ from unity
by demanding consistency between the numerical best fit expression for ${}^7 Li$ abundance~\cite{Steig2}
\be
y_{Li_p}=4.82\left(1\pm0.1\right)\left[\frac{\eta_{10}-3\left(Z-1\right)}{6}\right]^2
\ee
and observational bounds. In this case one has~\cite{Bhatta,Ghos,Fields}
\be
\delta Z\lesssim\mathcal{O}(10^{-1})\,.
\ee
Notice that this constraint is one order higher than
the corresponding value in Eq.~\eqref{ZbHe}. 

Thus, from Eq.~\eqref{Zbetabis} we obtain
\be
\label{bli}
\beta\lesssim\mathcal{O}(10^{82})\,,
\ee
for $T\simeq10\,\mathrm{MeV}$
and
\be
\label{bli2}
\beta\lesssim\mathcal{O}(10^{90})\,,
\ee
for $T\simeq0.1\,\mathrm{MeV}$. Also in this case, 
the approximation $\epsilon\ll1$ is satisfied, being
$\epsilon\sim\mathcal{O}(10^{-2})$.

As predictable, the overlap between 
the bound on $\beta$ from ${}^7 Li$ abundance on one hand
and ${}^4 He$, $D$ abundances on the other
is only partial, 
though non-vanishing. This discloses the possibility 
that the ${}^7 Li$ puzzle might be successfully addressed 
within the framework of GUP-modified Cosmology for a suitable
choice of the GUP parameter. 
Investigation along this direction requires further attention
and will be developed elsewhere.

\section{Discussion and Conclusions}
\label{DC}
Merging General Relativity and Quantum Theory
is one of the hottest topics in modern theoretical physics. 
A phenomenological approach to endow Quantum Mechanics
with gravity effects is to modify the Heisenberg Uncertainty
Principle in such a way as to reproduce a 
minimal observable length at Planck scale - Generalized Uncertainty Principle.
Although the natural domain of GUP is high-energy physics, 
the best - and, for the time being, unique - arena to quantify
the magnitude of GUP corrections is low-energy regime.
In this vein, it should be understood the large 
number of attempts to constrain the GUP deformation parameter
via optomechanical/interferometry experiments on one hand
and gravitational/cosmological measurements on the other (see~\cite{Hosse,ScardRev} for a review). 

Starting from the well-established connection
between the first law of thermodynamics and the cosmological Friedmann equations, in this work we have investigated the implications
of GUP on Big Bang Nucleosynthesis and the related abundances of 
primordial light elements. We emphasize
that GUP enters the Friedmann equations through a non-trivial modification
of the entropy area law (see Eq.~\eqref{mod}), which 
in turn affects the standard density/temperature dependence 
of Hubble constant and of its time derivative. GUP-corrected
Friedmann equations are given in Eqs.~\eqref{HT} and~\eqref{SGFM}
to the leading order in the deformation parameter.

Consistency with observational data on $i)$
variations of the freeze-out temperature $T_f$ and $ii)$
primordial abundances of $^{4}He$, $D$ and $^{7}Li$
has allowed us to infer various constraints on the GUP parameter $\beta$, 
the most stringent being $\beta\lesssim\mathcal{O}(10^{81})$
derived from the analysis of the $^{4}He$ and $D$ abundances. 
It is worth noticing that such bound fits
with those found in~\cite{Kouwn,Giardino} from similar
cosmological studies, although it is
less stringent than constraints inferred via gravitational or 
quantum experiments. This somehow indicates
the negligible, though non-vanishing, 
r\^ole of the GUP on cosmological scales.
In this sense, it would be interesting to study 
implications of the Extended Uncertainty Principle (EUP)~\cite{Bolen,Park,Mignemi,Gine}, which naturally emerges
in spacetime with a maximal length (horizon-like) scale, 
such as (anti)-de Sitter background. Besides 
this aspect, another important result of this work
is the possibility that the  $^{7}Li$ problem could
be solved in the framework of modified GUP Cosmology.

A further direction to explore is the study of effects
of other GUP formulations on BBN cosmological 
model. Indeed, as argued at the end
of the previous Section, higher-order GUP corrections terms might
be relevant, particularly in the study of the $^{7}Li$
problem. 

Finally,  we mention that a similar analysis has been carried out in~\cite{Ghos} in the context of non-extensive Tsallis Cosmology, 
which is a generalization of the ordinary Cosmology
based on Tsallis non-additive definition of horizon entropy~\cite{Tsallis}. 
In light of this extension, it is worth investigating
whether a connection between GUP and Tsallis frameworks
can be established, so as to map the GUP parameter
and Tsallis non-extensivity index into each other. 
Work along these and other directions is presently under active
consideration and will be presented in future works.


\smallskip

\begin{thebibliography}{0}

\section*{References}

\bibitem{Amati}
D. Amati, M. Ciafaloni, G. Veneziano, Phys. Lett. B \textbf{197}, 81
(1987).

\bibitem{Konishi}
K. Konishi, G. Paffuti, P. Provero, Phys. Lett. B \textbf{234}, 276 (1990).

\bibitem{Maggiore}
M. Maggiore, Phys. Lett. \textbf{B} 319, 83 (1993).

\bibitem{Kempf}
A. Kempf, G. Mangano, R.B. Mann, Phys. Rev. D \textbf{52}, 1108 (1995).

\bibitem{Scard}
 F. Scardigli, Phys. Lett. B \textbf{452}, 39 (1999). 

\bibitem{Capozzi}
S. Capozziello, G. Lambiase, G. Scarpetta, Int. J. Theor. Phys. \textbf{39},
15 (2000).
 
\bibitem{Adler}
R.~J.~Adler, P.~Chen and D.~I.~Santiago,
%``The Generalized uncertainty principle and black hole remnants,''
Gen. Rel. Grav. \textbf{33}, 2101 (2001).

\bibitem{Mague}
J. Magueijo and L. Smolin, Phys. Rev. Lett. \textbf{88}, 190403 (2002).

\bibitem{Garay}
L. J. Garay, Int. J. Mod. Phys. A \textbf{10}, 145 (1995).

\bibitem{Medved}
A. J. M. Medved and E. C. Vagenas, Phys. Rev. D \textbf{70}, 124021 (2004).

\bibitem{AmeCamel}
G. Amelino-Camelia, M. Arzano and A. Procaccini, Phys. Rev. D \textbf{70}, 107501 (2004).

  \bibitem{Brau}
 F.~Brau,
  %``Minimal length uncertainty relation and hydrogen atom,''
  J.\ Phys.\ A {\bf 32} 7691 (1999).
%
 \bibitem{Das}
S.~Das and E.~C.~Vagenas,
%``Universality of Quantum Gravity Corrections,''
Phys. Rev. Lett. \textbf{101}, 221301 (2008).
%
\bibitem{Pedram}
P.~Pedram, K.~Nozari and S.~H.~Taheri,
  %``The effects of minimal length and maximal momentum on the transition rate of ultra cold neutrons in gravitational field,''
  JHEP {\bf 1103}, 093 (2011).
%
\bibitem{ScardCas}
F.~Scardigli and R.~Casadio,
  %``Gravitational tests of the Generalized Uncertainty Principle,''
   Eur.\ Phys.\ J.\ C {\bf 75}, 425 (2015).
%
\bibitem{QC}
F.~Scardigli, G.~Lambiase and E.~Vagenas,
  %``GUP parameter from quantum corrections to the Newtonian potential,''
  Phys.\ Lett.\ B {\bf 767}, 242 (2017).
%  %
\bibitem{Petroz}
G.~G.~Luciano and L.~Petruzziello,
%``GUP parameter from Maximal Acceleration,''
Eur. Phys. J. C \textbf{79},  283 (2019).
%
\bibitem{CGgrav}
L.~Buoninfante, G.~G.~Luciano and L.~Petruzziello,
%``Generalized Uncertainty Principle and Corpuscular Gravity,''
Eur. Phys. J. C \textbf{79}, 663 (2019).
%
\bibitem{Buoninf}
L.~Buoninfante, G.~Lambiase, G.~G.~Luciano and L.~Petruzziello,
%``Phenomenology of GUP stars,''
Eur. Phys. J. C \textbf{80}, 853 (2020).
%
\bibitem{AliTest}
A.~F.~Ali, S.~Das and E.~C.~Vagenas,
%``A proposal for testing Quantum Gravity in the lab,''
Phys. Rev. D \textbf{84}, 044013 (2011).
%
\bibitem{Bruk}
 I.~Pikovski, M.~R.~Vanner, M.~Aspelmeyer, M.~S.~Kim and C.~Brukner,
%``Probing Planck-scale physics with quantum optics,''
Nature Phys. \textbf{8}, 393 (2012).
%
 \bibitem{GravBar}
F.~Marin, F.~Marino, M.~Bonaldi, M.~Cerdonio, L.~Conti, P.~Falferi, R.~Mezzena, A.~Ortolan, G.~A.~Prodi and L.~Taffarello, \textit{et al.}
%``Gravitational bar detectors set limits to Planck-scale physics on macroscopic variables,''
Nature Phys. \textbf{9}, 71 (2013).
%
\bibitem{Bawaj}
M.~Bawaj, C.~Biancofiore, M.~Bonaldi, F.~Bonfigli, A.~Borrielli, G.~Di Giuseppe, L.~Marconi, F.~Marino, R.~Natali and A.~Pontin, \textit{et al.}
%``Probing deformed commutators with macroscopic harmonic oscillators,''
Nature Commun. \textbf{6}, 7503 (2015). 
%
 \bibitem{BossoLigo}
P.~Bosso, S.~Das and R.~B.~Mann,
%``Potential tests of the Generalized Uncertainty Principle in the advanced LIGO experiment,''
Phys. Lett. B \textbf{785},  498 (2018).
%
\bibitem{Pendu}
P.~A.~Bushev, J.~Bourhill, M.~Goryachev, N.~Kukharchyk, E.~Ivanov, S.~Galliou, M.~E.~Tobar and S.~Danilishin,
%``Testing the generalized uncertainty principle with macroscopic mechanical oscillators and pendulums,''
Phys. Rev. D \textbf{100}, 066020 (2019).
%
\bibitem{LucLuc}
G.~G.~Luciano and L.~Petruzziello,
%``Generalized uncertainty principle and its implications on geometric phases in quantum mechanics,''
Eur. Phys. J. Plus \textbf{136}, 179 (2021). 
%

\bibitem{LucianoCas}
F.~Scardigli, M.~Blasone, G.~Luciano and R.~Casadio,
%``Modified Unruh effect from Generalized Uncertainty Principle,''
Eur. Phys. J. C \textbf{78}, 728 (2018).

\bibitem{Kana}
 T. Kanazawa, G. Lambiase, G. Vilasi, A. Yoshioka, Eur. Phys. J. C
\textbf{79}, 95 (2019).

\bibitem{BossoLuc}
P.~Bosso and G.~G.~Luciano,
%``Generalized Uncertainty Principle: from the harmonic oscillator to a QFT toy model,''
Eur. Phys. J. C \textbf{81}, 982 (2021). 

\bibitem{JizbaKl}
P.~Jizba, H.~Kleinert and F.~Scardigli, Phys.\ Rev.\ D {\bf 81}, 084030 (2010).

\bibitem{Ong}
Y.~C.~Ong,
%``Generalized Uncertainty Principle, Black Holes, and White Dwarfs: A Tale of Two Infinities,''
JCAP \textbf{09}, 015 (2018).

\bibitem{BLPS}
L.~Buoninfante, G.~G.~Luciano, L.~Petruzziello and F.~Scardigli,
%``Bekenstein bound and uncertainty relations,''
[arXiv:2009.12530 [hep-th]].

\bibitem{Anacleto}
M. Anacleto, F. Brito, E. Passos, Phys. Lett. B \textbf{749},
181 (2015).

\bibitem{Jacob}
T. Jacobson, Phys. Rev. Lett. \textbf{75}, 1260 (1995).

\bibitem{Wang}
B. Wang, E. Abdalla, R.K. Su, Phys. Lett. B \textbf{503}, 394 (2001).

\bibitem{Frolov}
A. V. Frolov and L. Kofman, JCAP 0305, \textbf{009} (2003).

\bibitem{Padmanabhan1}
T. Padmanabhan, Phys. Rep. \textbf{406}, 49 (2005). 

\bibitem{Eling}
C.~Eling, R.~Guedens and T.~Jacobson,
%``Non-equilibrium thermodynamics of spacetime,''
Phys. Rev. Lett. \textbf{96}, 121301 (2006). 

\bibitem{Akbar1}
M.~Akbar and R.~G.~Cai,
%``Friedmann equations of FRW universe in scalar-tensor gravity, f(R) gravity and first law of thermodynamics,''
Phys. Lett. B \textbf{635}, 7 (2006). 

\bibitem{Cai}
R.-G. Cai, S.P. Kim, JHEP \textbf{02}, 050 (2005).

\bibitem{CaiCao}
R.-G. Cai, L.-M. Cao, Y.-P. Hu, JHEP \textbf{08}, 090 (2008).

\bibitem{Zhu}
T. Zhu, J.-R. Ren, M.-F. Li, Phys. Lett. B \textbf{674}, 204 (2009). 

\bibitem{LambBar}
S.~Das, M.~Fridman, G.~Lambiase and E.~C.~Vagenas,
%``Baryon Asymmetry from the Generalized Uncertainty Principle,''
[arXiv:2107.02077 [gr-qc]].

\bibitem{Giardino}
S.~Giardino and V.~Salzano,
%``Cosmological constraints on the Generalized Uncertainty Principle from modified Friedmann equations,''
Eur. Phys. J. C \textbf{81}, 110 (2021). 

\bibitem{Kouwn}
S.~Kouwn,
%``Implications of Minimum and Maximum Length Scales in Cosmology,''
Phys. Dark Univ. \textbf{21}, 76 (2018). 

\bibitem{Ghosh}
S. Ghosh, Class. Quantum Gravity \textbf{31}, 025025 (2014).

\bibitem{Gao}
D. Gao, J.Wang, M. Zhan, Phys. Rev.A \textbf{95}, 042106 (2017).

\bibitem{Luciano:2021cna}
G.~G.~Luciano and L.~Petruzziello,
%``Generalized uncertainty principle and its implications on geometric phases in quantum mechanics,''
Eur. Phys. J. Plus \textbf{136}, 179 (2021). 

\bibitem{ColdGao}
D. Gao, M. Zhan, Phys. Rev. A \textbf{94}, 013607 (2016).

\bibitem{Feng}
Z.-W. Feng, S.-Z. Yang, H.-L. Li, X.-T. Zu, Phys. Lett. B
\textbf{768}, 81 (2017). 

\bibitem{Shap}
\"O.~\"Okc\"u and E.~Aydiner,
%``Observational tests of the generalized uncertainty principle: Shapiro time delay, gravitational redshift, and geodetic precession,''
Nucl. Phys. B \textbf{964}, 115324 (2021).

\bibitem{Jusufi}
K.~Jusufi, M.~Azreg-A\"\i{}nou, M.~Jamil and T.~Zhu,
%``Constraining the Generalized Uncertainty Principle Through Black Hole Shadow and Quasiperiodic Oscillations,''
[arXiv:2008.09115 [gr-qc]].

\bibitem{Neves}
J.~C.~S.~Neves,
%``Upper bound on the GUP parameter using the black hole shadow,''
Eur. Phys. J. C \textbf{80}, 343 (2020).

\bibitem{Kolb}
E.W. Kolb, M.S. Turner, \emph{The Early Universe}, Addison
Wesley Publishing Company (1989).

\bibitem{Bern}
J. Bernstein, L.S. Brown, G. Feinberg, Rev. Mod. Phys.
\textbf{61}, 25 (1989).

\bibitem{PDG}
P.~A.~Zyla \textit{et al.} [Particle Data Group],
%``Review of Particle Physics,''
PTEP \textbf{2020}, 083C01 (2020).

\bibitem{Ghos}
A.~Ghoshal and G.~Lambiase,
%``Constraints on Tsallis Cosmology from Big Bang Nucleosynthesis and Dark Matter Freeze-out,''
[arXiv:2104.11296 [astro-ph.CO]].

\bibitem{AliGUP}
A.~Awad and A.~F.~Ali,
%``Minimal Length, Friedmann Equations and Maximum Density,''
JHEP \textbf{06}, 093 (2014). 

\bibitem{LambJCAP}
G.~Lambiase,
%``Constraints on massive gravity theory from big bang nucleosynthesis,''
JCAP \textbf{10}, 028 (2012).

\bibitem{CapLamb}
S.~Capozziello, G.~Lambiase and E.~N.~Saridakis,
%``Constraining f(T) teleparallel gravity by Big Bang Nucleosynthesis,''
Eur. Phys. J. C \textbf{77}, 576 (2017).

\bibitem{Aver}
E.~Aver, K.~A.~Olive and E.~D.~Skillman,
%``The effects of He I \ensuremath{\lambda}10830 on helium abundance determinations,''
JCAP \textbf{07}, 011 (2015). 

\bibitem{Boran}
S.~Boran and E.~O.~Kahya,
%``Testing a Dilaton Gravity Model using Nucleosynthesis,''
Adv. High Energy Phys. \textbf{2014}, 282675 (2014). 

\bibitem{Bhatta}
S. Bhattacharjee, P.K. Sahoo, Eur. Phys. J. Plus \textbf{135},
350 (2020).

\bibitem{Kneller}
J.P. Kneller, G. Steigman, New J. Phys. \textbf{6}, 117 (2004).

\bibitem{Steig}
G. Steigman, Annu. Rev. Nucl. Part. Sci. \textbf{57}, 463 (2007). 

\bibitem{Fields}
B.~D.~Fields, K.~A.~Olive, T.~H.~Yeh and C.~Young,
%``Big-Bang Nucleosynthesis after Planck,''
JCAP \textbf{03}, 010 (2020)
[erratum: JCAP \textbf{11}, E02 (2020)]. 

\bibitem{Steig2}
G. Steigman, Adv. High Energy Phys. \textbf{2012}, 268321 (2012). 

\bibitem{Fields2}
B.D. Fields, Annu. Rev. Nucl. Part. Sci. \textbf{61}, 47 (2011).

\bibitem{Hosse}
S.~Hossenfelder,
%``Minimal Length Scale Scenarios for Quantum Gravity,''
Living Rev. Rel. \textbf{16}, 2 (2013). 

\bibitem{ScardRev}
F.~Scardigli,
%``The deformation parameter of the generalized uncertainty principle,''
J. Phys. Conf. Ser. \textbf{1275}, 012004 (2019). 

\bibitem{Bolen}
B.~Bolen and M.~Cavaglia,
%``(Anti-)de Sitter black hole thermodynamics and the generalized uncertainty principle,''
Gen. Rel. Grav. \textbf{37}, 1255 (2005). 

\bibitem{Park}
M.~i.~Park,
%``The Generalized Uncertainty Principle in (A)dS Space and the Modification of Hawking Temperature from the Minimal Length,''
Phys. Lett. B \textbf{659}, 698 (2008). 

\bibitem{Mignemi}
S.~Mignemi,
%``Extended uncertainty principle and the geometry of (anti)-de Sitter space,''
Mod. Phys. Lett. A \textbf{25}, 1697 (2010).

\bibitem{Gine}
J.~Gin\'e and G.~G.~Luciano,
%``Modified inertia from extended uncertainty principle(s) and its relation to MoND,''
Eur. Phys. J. C \textbf{80}, 1039 (2020). 

\bibitem{Tsallis}
C. Tsallis, J. Statist. Phys. \textbf{52}, 479 (1988).

\end{thebibliography}
\end{document}